%% file: bs2etapxss_post_proofs.tex
\documentclass[aps,prd,reprint,superscriptaddress]{revtex4-2}

\usepackage{relsize}
\def\babar{\mbox{\slshape B\kern-0.1em{\smaller A}\kern-0.1em
		B\kern-0.1em{\smaller A\kern-0.2em R \space}}}

\usepackage[utf8]{inputenc}
\usepackage{graphicx}
\usepackage{lineno}
\usepackage{amsmath}
\usepackage{graphicx}
\usepackage{subfigure}

\begin{document}
	
\preprint{\vbox{ \hbox{   }
		\hbox{Belle Preprint 2021-01}
		\hbox{KEK Preprint 2020-40}
}}

\title{Search for $B_{s}^{0} \rightarrow \eta^{\prime} X_{s\bar{s}}$ at Belle using a semi-inclusive method}

\input{pub574.tex}

\date{\today}

\begin{abstract}
\clearpage
We report the first search for the penguin-dominated process $B_{s}^{0} \rightarrow \eta^{\prime} X_{s\bar{s}}$ using a semi-inclusive method.  A 121.4 $\mathrm{fb}^{-1}$ integrated luminosity $\Upsilon(5S)$ data set collected by the Belle experiment, at the KEKB asymmetric-energy $e^+e^-$ collider, is used.  We observe no statistically significant signal and, including all uncertainties, we set a 90\% confidence level upper limit on the partial branching fraction at 1.4 $\times$ 10$^{-3}$ for $M(X_{s\bar{s}})$ $\leq$ 2.4 GeV/$c^{2}$.
\end{abstract}
%X_{s\bar{s}}
%\keywords{}

\maketitle

%\section{}
%\subsection{}
%\subsubsection{}

% If in two-column mode, this environment will change to single-column
% format so that long equations can be displayed. Use
%\begin{widetext}
%\end{widetext}

%%%%%%%%%%%%%%%%%%%%% begin text %%%%%%%%%%%%%%%%%%%%%%%%%

The study of the decay of $B$ mesons --- bound states of a $b$ antiquark and either a $u$, $d$, $s$, or $c$ quark --- has been fruitful for the interrogation of rare processes, elucidating the strong and weak interactions of the Standard Model (SM) of particle physics.  According to the SM flavor-changing neutral currents are forbidden in $B$ decays at leading-order, but may effectively occur at higher-order in ``penguin" $\Delta B = 1$ processes, where $B$ is the beauty quantum number \cite{Bevan_2014}. 

The CLEO collaboration measured a larger than expected branching fraction (BF) for the charmless decay (decays whose primary decay products lack a charm quark) $B\rightarrow \eta^{\prime} X_{s}$ as $\mathcal{B}(B \rightarrow \eta^{\prime} X_{s})$ = [4.6 $\pm$ 1.1 (stat.) $\pm$ 0.4 (syst.) $\pm$ 0.5 (bkg.)]$\times$ 10$^{-4}$, with $M(X_{s}) < $ 2.35 GeV/$c^{2}$, where the third uncertainty is due to the background subtraction \cite{PhysRevLett.81.1786,Bonvicini_2003}.  \babar measured $\mathcal{B}(B \rightarrow \eta^{\prime} X_{s})$ = [3.9 $\pm$ 0.8 (stat.) $\pm$ 0.5 (syst.) $\pm$ 0.8 (model)] $\times$ 10$^{-4}$, for the same $M(X_{s})$ requirement \cite{PhysRevLett.93.061801}.  Here, ``model" refers to the fragmentation uncertainty of the $X_{s}$.  Belle previously measured the BF for the  related process $B \rightarrow \eta X_{s}$ as $\mathcal{B}$($B \rightarrow \eta X_{s}$) = [26.1 $\pm$ 3.0 (stat.) $^{+1.9}_{-2.1}$ (syst.) $^{+4.0}_{-7.1}$ (model)] $\times 10^{-5}$ \cite{Nishimura:2009ae}.  

While the $\eta^{\prime}$ meson itself is interesting \cite{PhysRevD.97.054508} as its mass is higher than is expected from symmetry considerations, it is the unexpected BF enhancement seen in the $B \rightarrow \eta^{\prime} X_{s}$ measurements that has generated considerable interest.  In Ref. \cite{DATTA1998369}, for example, the predicted BF for a four-quark SM prediction for $B \rightarrow \eta^{\prime} X_{s}$ is 1.3 $\times 10^{-4}$.  Explanations for this apparent enhancement focus on processes such as the $b \rightarrow sg$ transition, which is modified to an anomalous $b \rightarrow sg^{*}$ process, where $g^{*} \rightarrow g \eta^{\prime}$, with the gluon coupling to the $\eta^{\prime}$ singlet \cite{Atwood:1997bn,petrovkagan1997,FRITZSCH199783,PhysRevLett.80.434,alipark2001,HE1999123,PhysRevD.63.054027}.  Hence, glueball coupling may provide an explanation for these decays involving the $\eta^{\prime}.$

Inclusive $b \rightarrow sg$ processes have not yet been investigated using the $B_{s}^{0}$ meson.  We report the first search for the decay $B_{s}^{0} \rightarrow \eta^{\prime} X_{s\bar{s}}$ using a semi-inclusive method \cite{Shawn:2344} with data collected at the $\Upsilon(5S)$ resonance by the Belle detector at the KEKB asymmetric-energy $e^{+}e^{-}$ collider in Japan \cite{10.1093/ptep/pts102}.

\begin{figure}
	\centering     %%% not \center
	\subfigure[QCD Penguin]{\label{fig:a}\includegraphics[width=35mm]{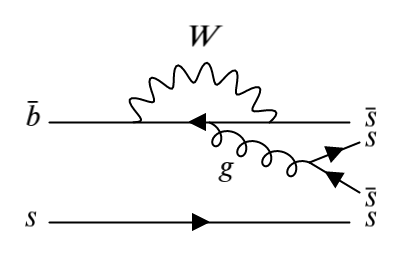}}
	\qquad
	\subfigure[QCD Penguin]{\label{fig:b}\includegraphics[width=35mm]{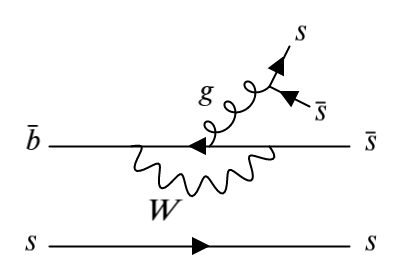}}
	\qquad
	\subfigure[$g-\eta^{\prime}$ Coupling]{\label{fig:c}\includegraphics[width=35mm]{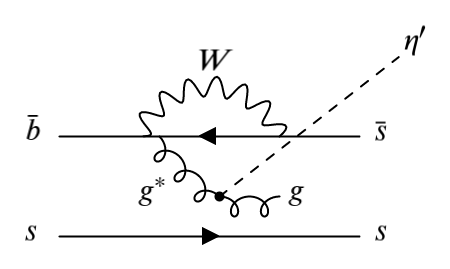}}
	\qquad
	\subfigure[Color-Suppressed Tree]{\label{fig:d}\includegraphics[width=35mm]{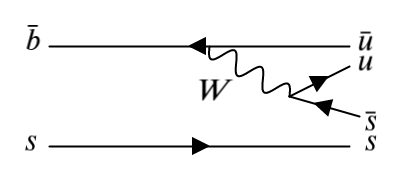}}
	\qquad
	\subfigure[Electroweak Penguin]{\label{fig:e}\includegraphics[width=35mm]{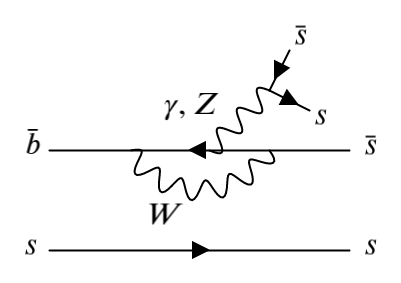}}
	\qquad
	\subfigure[Color-Suppressed Electroweak Penguin]{\label{fig:f}\includegraphics[width=35mm]{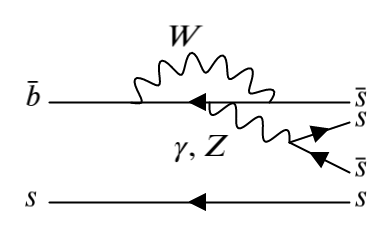}}
	\caption{Lowest-order diagrams contributing to $B_{s}^{0} \rightarrow \eta^{\prime} X_{s\bar{s}}$.}
	\label{fig:diagrams}
\end{figure}

To lowest order, the amplitude for $B_{s}^{0} \rightarrow \eta^{\prime} X_{s\bar{s}}$ contains contributions from QCD penguin diagrams \footnote{In Fig. \ref{fig:diagrams}(b), there is a soft gluon in the hadronization process, which is required to conserve color but is conventionally not shown in the Feynman diagram \cite{DATTA2020}}, the anomalous $g \eta^{\prime}$ coupling, the tree-level color-suppressed $b \rightarrow u$ diagram, and the $b \rightarrow s (\gamma,Z)$ electroweak penguin diagrams, shown in Fig. \ref{fig:diagrams}.  Contributions from penguin annihilation diagrams are typically omitted as they are suppressed by a factor of $\Lambda_{\mathrm{QCD}}/m_{b}$, where $\Lambda_{\mathrm{QCD}}$ is the quantum chromodynamic scale and $m_{b}$ is the mass of the beauty quark \cite{DATTA2020}.

The Belle detector is a large-solid-angle magnetic
spectrometer that consists of a silicon vertex detector (SVD),
a 50-layer central drift chamber (CDC), an array of
aerogel threshold Cherenkov counters (ACC),
a barrel-like arrangement of time-of-flight
scintillation counters (TOF), and an electromagnetic calorimeter
comprised of CsI(Tl) crystals (ECL) located inside 
a superconducting solenoid coil that provides a 1.5~T
magnetic field.  An iron flux-return located outside of
the coil is instrumented to detect $K_L^0$ mesons and to identify
muons.  For the $\Upsilon(5S)$ data sample, Belle used a 1.5 cm radius beampipe, a 4-layer SVD, and a small-inner-cell CDC \cite{10.1093/ptep/pts072}.

We use the 121.4 $\mathrm{fb}^{-1}$ data sample recorded by Belle, taken at the center-of-mass (CM) energy $\sqrt{s}=10.866$ $\mathrm{GeV}$, which corresponds to the $\Upsilon(5S)$ resonance.  The $\Upsilon(5S)$ decays to $B_{s}^{0}$ pairs with a branching fraction of 0.172 $\pm$ 0.030  and of this fraction the $\Upsilon(5S)$ has three channels for the $B_{s}^{0}$ decays: $\Upsilon(5S)$ $\rightarrow B_{s}^{0*} \bar{B}_{s}^{0*}$, $\Upsilon(5S)$ $\rightarrow B_{s}^{0} \bar{B}_{s}^{0*}$ and $B_{s}^{0*} \bar{B}_{s}^{0}$, and $\Upsilon(5S)$ $\rightarrow B_{s}^{0} \bar{B}_{s}^{0}$.  The rates are 87.0\%, 7.3\%, and 5.7\%, respectively \cite{PhysRevD.87.031101}.  This corresponds to (7.11 $\pm$ 1.30) $\times$ 10$^{6}$ $B_{s}^{0}\bar{B}_{s}^{0}$ pairs, the world's largest $\Upsilon(5S)$ sample in $e^{+}e^{-}$ collisions.  A blind analysis is performed, whereby the selection criteria are first optimized on Monte Carlo (MC) simulations before being applied to the data.  A signal MC sample for $B_{s}^{0} \rightarrow \eta^{\prime} X_{s\bar{s}}$ is generated using EvtGen \cite{Lange:2001uf} and the detector response is simulated using GEANT3 \cite{Brun:1119728}, with PHOTOS describing final-state radiation \cite{BARBERIO1991115}.  The MC-generated mass of the $X_{s\bar{s}}$ system is bounded below by the two-(charged) kaon mass 0.987 GeV/$c^{2}$ and has an upper bound of 3.0 GeV/$c^{2}$.  The $X_{s\bar{s}}$ mass is generated as a flat distribution and is fragmented by $\textsc{Pythia}$ 6 \cite{Sj_strand_2006}.  The flat distribution reduces model dependence and allows for an analysis that does not depend on the $X_{s\bar{s}}$ mass distribution.

The $B_{s}^{0}$ ($\bar{b}s$) and $\bar{B}_{s}^{0}$ ($b\bar{s}$) candidates are reconstructed using a semi-inclusive method in which the $X_{s\bar{s}}$ is reconstructed as a system of two kaons, either $K^{+}K^{-}$ or $K^{\pm}K_{S}^{0} (\rightarrow \pi^{+} \pi^{-})$, and up to four pions with at most one $\pi^{0}$, where the $\pi^{0}$ decays via the channel $\pi^{0} \rightarrow \gamma \gamma$.  The $\eta^{\prime}$ is reconstructed in the channel $\eta^{\prime} \rightarrow \eta (\rightarrow \gamma \gamma) \pi^{+} \pi^{-}$.  The experimental signature is divided into two classes of decay modes: without ($B_{s}^{0} \rightarrow \eta^{\prime} K^{+} K^{-} + n\pi$) and with ($B_{s}^{0} \rightarrow \eta^{\prime} K^{\pm} K_{S}^{0} + n\pi$) a $K_{S}^{0}$.  These classes are analyzed separately, with the weighted average BFs taken at the end.  Charge-conjugate decays are included unless explicitly stated otherwise.

Charged particle tracks are required to satisfy loose impact parameter requirements to remove mismeasured tracks \cite{Shawn:2344}, and have transverse momenta $p_{\rm T}$ greater than 50 MeV/$c$.  Separation of the charged kaons and charged pions is provided by the CDC \cite{HIRANO2000294}, ACC \cite{IIJIMA2000321}, and the TOF \cite{KICHIMI2000315} systems.  Information from these subdetectors is combined to form a likelihood ratio for the charged kaon hypothesis: $P_{K^{\pm}}$ = $L_{K^{\pm}}$/($L_{K^{\pm}}$ + $L_{\pi^{\pm}}$).  For this analysis, the selections $P_{K^{\pm}}$ $>$ 0.6 for $K^{\pm}$ and $P_{K^{\pm}}$ $<$ 0.6 for $\pi^{\pm}$ are applied.  The efficiency to correctly identify a pion (kaon) is 98\% (88)\%, with a misidentification rate of 4\% (12)\% \cite{Nishimura:2009ae}.

The $\pi^{0}$ candidate mass range is $M(\gamma \gamma)$ $\in$ [0.089, 0.180] GeV/$c^{2}$ ($\pm$5$\sigma$ window).  The $\pi^{0}$ candidates are kinematically constrained to the nominal mass \cite{Tanabashi:2018oca}.  In the ECL, the photons constituting the $\pi^{0}$ are required to have energies greater than 50 MeV in the barrel region, greater than 100 MeV in the endcaps, and the ratio of their energy depositions in a 3$\times$3 ECL crystal array to that in a 5$\times$5 crystal array around the central crystal, is required to be greater than 0.9.  To further reduce combinatorial background, a requirement on the $\pi^{0}$ laboratory-frame momentum to be greater than 0.2 GeV/$c$ is imposed.

The $\eta$ is reconstructed in a two-photon asymmetric invariant mass window $M_{\eta} \in$ [0.476, 0.617] GeV/$c^{2}$ (4.5$\sigma_{L}$, 9.2$\sigma_{R}$, from signal MC samples, after all final selections are applied), where $L$ and $R$ refer to the left and right sides of the mean of the mass distribution.  The asymmetry is due to energy leakage in the ECL, causing the $\eta$ mass distribution to be asymmetric.  Each photon is required to have $E_{\gamma} >$ 0.1 GeV.  A requirement on the photon-energy asymmetry ratio $|E_{\gamma1} - E_{\gamma2}|/(E_{\gamma1} + E_{\gamma2}) <$ 0.6 is applied to further suppress the background.  The $\eta^{\prime}$ mesons are reconstructed in a maximally efficient mass window $M_{\eta^{\prime}} \in$ [0.933, 0.982] GeV/$c^{2}$ (approximately $\pm$7.0$\sigma$, from signal MC samples, after all final selections are applied).  The $\eta$ and $\eta^{\prime}$ masses are kinematically fit to the world average \cite{Tanabashi:2018oca}.  The mass range of the $K_{S}^{0}$ is $M_{K_{S}^{0}} \in$ [0.487, 0.508] GeV/$c^{2}$ ($\pm$3$\sigma$ window).

The $X_{s\bar{s}}$ system is reconstructed as a system of kaons and pions, which is in turn combined with the $\eta^{\prime}$ to form $B_{s}$ candidates.  Two variables important in extracting the signal are the energy difference $\Delta E$, defined as $\Delta E = E_{B_{s}} -  E_{\rm beam}$ and the beam-energy-constrained mass, defined as $M_{\mathrm{bc}} = \sqrt{E_{\rm beam}^2/c^{4} - p_{B_{s}}^{2}/c^{2}}$, where $E_{\rm beam} = \sqrt{s}/2$, $E_{B_{s}}$ is the energy of the $B_{s}$, and $p_{B_{s}}$ is the magnitude of the $B_{s}$ three-momentum in the CM frame of the colliding $e^{+}e^{-}$ beams.

The dominant nonpeaking background is from continuum with others coming from generic $B_{s}^{0(*)}\bar{B}_{s}^{0(*)}$ and $B\bar{B}X$ decays.  An initial reduction in continuum background ($e^{+}e^{-} \rightarrow q\bar{q}$, $q$ = $u$, $d$, $s$, $c$) is done with a selection on the ratio of the second to the zeroth order Fox-Wolfram moments $R_{2} \leq$ 0.6 \cite{PhysRevLett.41.1581}.  A neural network (NN), NeuroBayes \cite{FEINDT2006190}, is used to further suppress continuum background, with other backgrounds being reduced as well.  The NN is trained to primarily discriminate between event topologies using event shape variables \cite{PhysRevLett.91.261801}.  Signal events have a spherical topology, while continuum background events are jetlike.  The NN is trained using these variables on independent signal and continuum background MC simulations.  The NN output variable $O_{\rm NN}$ describes, effectively, the probability that a $B_{s}^{0}$ candidate came from an event whose topology is spherical or jetlike.

To obtain a specific $O_{\rm NN}$ selection, the figure-of-merit (FOM) $S/\sqrt{S + B}$ is optimized as a function of $O_{\rm NN}$, where $S$ and $B$ are the fitted signal and background yields from an MC sample that is passed through the trained network.  This MC contains an approximately data-equivalent background and an enhanced signal.  This was done assuming $\mathcal{B}$($B_{s}^{0} \rightarrow \eta^{\prime} X_{s\bar{s}}$) = 2$\times$10$^{-4}$; this is 1.6 standard deviations below the \babar central value for $B \rightarrow \eta^{\prime} X_{s}$.  The value of $O_{\rm NN}$ corresponding to the maximum value of the FOM is selected.  Events having $O_{\rm NN}$ values below this selection are rejected.  Separate optimizations are done for $B_{s}^{0} \rightarrow \eta^{\prime} K^{+} K^{-} + n\pi$ and $B_{s}^{0} \rightarrow \eta^{\prime} K^{\pm} K_{S}^{0} + n\pi$, which have substantially different background levels and efficiencies.  The NN requirement reduces continuum background by more than 97\% in both cases, while preserving 39\% and 53\% of signal events for $B_{s}^{0} \rightarrow \eta^{\prime} K^{+} K^{-} + n\pi$ and $B_{s}^{0} \rightarrow \eta^{\prime} K^{\pm} K_{S}^{0} + n\pi$, respectively.  

After an initial requirement of $M_{\mathrm{bc}} > 5.30$ GeV/$c^{2}$, $|\Delta E| < 0.35$ GeV, and $M(X_{s\bar{s}})$ $\leq$ 2.4 GeV/$c^{2}$, and after all final selections are  applied, there are an average of 6.4 candidates per event for $B_{s}^{0} \rightarrow \eta^{\prime} K^{+} K^{-} + n\pi$ and 26.0 for $B_{s}^{0} \rightarrow \eta^{\prime} K^{\pm} K_{S}^{0} + n\pi$.    To select the best candidate per event, the candidate with the smallest $\chi^{2}$ given by $\chi^{2} = \chi^{2}_{\rm vtx}/ndf + (\Delta E - \mu_{\Delta E})^2/\sigma^2_{\Delta E}$ is selected, where $\Delta E$ is calculated on a candidate-by-candidate basis, and $\mu_{\Delta E}$ is the mean energy difference of the $\Delta E$ distribution, obtained through studies of signal MC of individual exclusive $B_{s}^{0} \rightarrow \eta^{\prime} X_{s\bar{s}}$ decay modes; $\sigma_{\Delta E}$ is the width of these distributions.  Here $\chi^{2}_{\rm vtx}/ndf$ is the reduced $\chi^{2}$ from a successful vertex fit of the primary charged daughter particles of the $X_{s\bar{s}}$.
From signal MC, the efficiency of the best candidate selection is 85.5\% for $B_{s}^{0} \rightarrow \eta^{\prime} K^{+} K^{-} + n\pi$ and 43.2\% for $B_{s}^{0} \rightarrow \eta^{\prime} K^{\pm} K_{S}^{0} + n\pi$, in the signal region.  The fraction of $B_{s}^{0}$ candidates passing best candidate selection that are correctly reconstructed is 94.0\% for $B_{s}^{0} \rightarrow \eta^{\prime} K^{+} K^{-} + n\pi$ and 60.4\% for $B_{s}^{0} \rightarrow \eta^{\prime} K^{\pm} K_{S}^{0} + n\pi$.  These numbers are obtained after all final selections are applied.

Other backgrounds were studied as sources of potential peaking background.  Due to the signal final state, it is difficult to have backgrounds that will be equivalent in topology and strangeness, and that are not highly suppressed.  However, one such unmeasured mode is $B_{s}^{0} \rightarrow \eta^{\prime} D_{s} \pi$.  Reconstruction efficiency is estimated using MC events and an expected number of peaking events is determined.  For $B_{s}^{0} \rightarrow \eta^{\prime} D_{s} \pi$ the BF is assumed to be similar to $B^{0} \rightarrow D^{-} \pi^{+} \rho^{0}$, for which the world average is [1.1 $\pm$ 1.0] $\times 10^{-3}$ \cite{Tanabashi:2018oca}.  After applying all final selections, the total number of expected peaking events is less than one.  There is a negligible amount of peaking background based on studies of $B_{(s)}^{0}\bar{B}_{(s)}^{0}$ MC samples.

The decay $B \rightarrow \eta^{\prime} K^{*0}$ can contribute to peaking background if the pion from $K^{*0} \rightarrow K^{-} \pi^{+}$ is misidentified.  The world average BF is [2.8 $\pm$ 0.6] $\times 10^{-6}$ \cite{Tanabashi:2018oca}.   From this and the pion misidentification rate, we expect the background contribution from this mode to be negligible.

The color-suppressed, tree-level process  $B_{s}^{0} \rightarrow \bar{D}^{0} \eta^{\prime}$, with $D^{0} \rightarrow K^{+}K^{-}$ could potentially contribute to the peaking background.  However, $B^{0} \rightarrow \bar{D}^{0} \eta^{\prime}$ has a measured BF of $\mathcal{B}(B^{0} \rightarrow \bar{D}^{0} \eta^{\prime})$ = [1.38 $\pm$ 0.16] $\times$ 10$^{-4}$.  The process $D^{0} \rightarrow K^{+}K^{-}$ is Cabibbo-suppressed and has a measured BF of $\mathcal{B}(D^{0} \rightarrow K^{+}K^{-})$ = [4.08 $\pm$ 0.06] $\times$ 10$^{-3}$ \cite{Tanabashi:2018oca}.  Assuming $SU(3)$ symmetry, we expect there to be less than one event from $B_{s}^{0} \rightarrow \bar{D}^{0} \eta^{\prime}$, for this analysis.

For signal extraction, fitting is done in 0.2 GeV/$c^{2}$ bins of $X_{s\bar{s}}$ mass, up to 2.4 GeV/$c^{2}$, using unbinned maximum-likelihood fits.  All submodes are combined for fitting.  Signal extraction is done by fitting the $M_{\mathrm{bc}}$ distribution in the region $M_{\mathrm{bc}} > 5.30$ GeV/$c^{2}$, $-$0.12 $\leq \Delta E \leq$ 0.05 GeV.

The $\Upsilon(5S)$ has three channels for $B_{s}^{0}$ decays: $\Upsilon(5S)$ $\rightarrow B_{s}^{0*} \bar{B}_{s}^{0*}$, $\Upsilon(5S)$ $\rightarrow B_{s}^{0} \bar{B}_{s}^{0*}$ and $B_{s}^{0*} \bar{B}_{s}^{0}$, and $\Upsilon(5S)$ $\rightarrow B_{s}^{0} \bar{B}_{s}^{0}$.  The corresponding rates are 87.0\%, 7.3\%, and 5.7\%, respectively \cite{PhysRevD.87.031101}.  The low-energy photon from $B_{s}^{0*} \rightarrow B_{s}^{0} \gamma$ is not reconstructed.  This has the effect of shifting the mean of the $\Delta E$ distribution to a value of approximately $-$50 MeV.  As a result, there are three signal peaks in the beam-energy-constrained mass distribution.

The signal in beam-energy-constrained mass is modeled as the sum of three Gaussian probability density functions (PDFs) that correspond to the three $\Upsilon(5S)$ decays described above.  Their shape parameters (means and widths of the signal Gaussians) are determined from a $B_{s}^{0} \rightarrow D_{s}^{-} \rho^{+}$ data control sample and are fixed in the fit to data.  The nonpeaking background fit component is an ARGUS PDF \cite{Albrecht:1990am} with a fixed shape parameter, determined from fits to $\Upsilon(5S)$ data NN sidebands.  The ARGUS endpoint is fixed at 5.434 GeV/$c^{2}$, the kinematic limit of $M_{\mathrm{bc}}$.  The full model is the sum of the signal and background PDFs, with the signal and background yields allowed to float.

The signal reconstruction efficiency, defined as $\epsilon_{i} = N^{\rm rec}_{i}/N^{\rm gen}_{i}$, is determined from fitting signal MC sample, in each $X_{s\bar{s}}$ mass bin $i$ after all selections are applied.  Here, $N^{\rm gen}_{i} = N^{B_{s}^{0} \rightarrow \eta^{\prime} K^{+} K^{-} + n\pi}_{i} + N^{B_{s}^{0} \rightarrow \eta^{\prime} K^{\pm} K^{0}_{S} + n\pi}_{i} + N^{\rm other}_{i}$, is the number of generated $B_{s}^{0}$ mesons in the signal MC sample.  The quantity $N^{\rm other}_{i}$ is the number of generated $B_{s}^{0}$ mesons that do not belong to either of the two classes of signal modes: $B_{s}^{0} \rightarrow \eta^{\prime} K^{+} K^{-} + n\pi$ and $B_{s}^{0} \rightarrow \eta^{\prime} K^{\pm} K^{0}_{S} + n\pi$ \footnote{The number of unreconstructed modes is discussed in the the appendix}.  The quantity $N^{\rm rec}_{i}$ is the number of events found from the Gaussian signal fit in the $i$th $X_{s\bar{s}}$ mass bin.

The BF is calculated as
$\mathcal{B}(B_{s}^{0} \rightarrow \eta^{\prime} X_{s\bar{s}})_{i} = N_{i}^{\mathrm{sig}}/[2 \times N_{B_{s}^{0(*)}\bar{B}_{s}^{0(*)}}\epsilon_{i}^{\prime}\mathcal{B}(\eta \rightarrow \gamma\gamma)\mathcal{B}(\eta^{\prime}  \rightarrow \pi^{+}\pi^{-}\eta)]$, where $i$ denotes the mass bins of $X_{s\bar{s}}$, the $\epsilon_{i}^{\prime}$ are the bin-by-bin MC signal reconstruction efficiencies $\epsilon_{i}$, corrected for data-MC discrepancies in NN selection, best candidate selection, particle identification, tracking efficiency, $\eta \rightarrow \gamma \gamma$ reconstruction, $\pi^{0} \rightarrow \gamma \gamma$ reconstruction, and $K^{0}_{S} \rightarrow \pi^{+} \pi^{-}$ reconstruction.  The quantity $N_{i}^{\mathrm{sig}}$ is the number of fitted signal events and the quantity $N_{B_{s}^{0(*)}\bar{B}_{s}^{0(*)}}$ is the total number of produced $B_{s}^{0}\bar{B}_{s}^{0}$ pairs.

\begin{figure}[!htb]
	\includegraphics[scale=0.1]{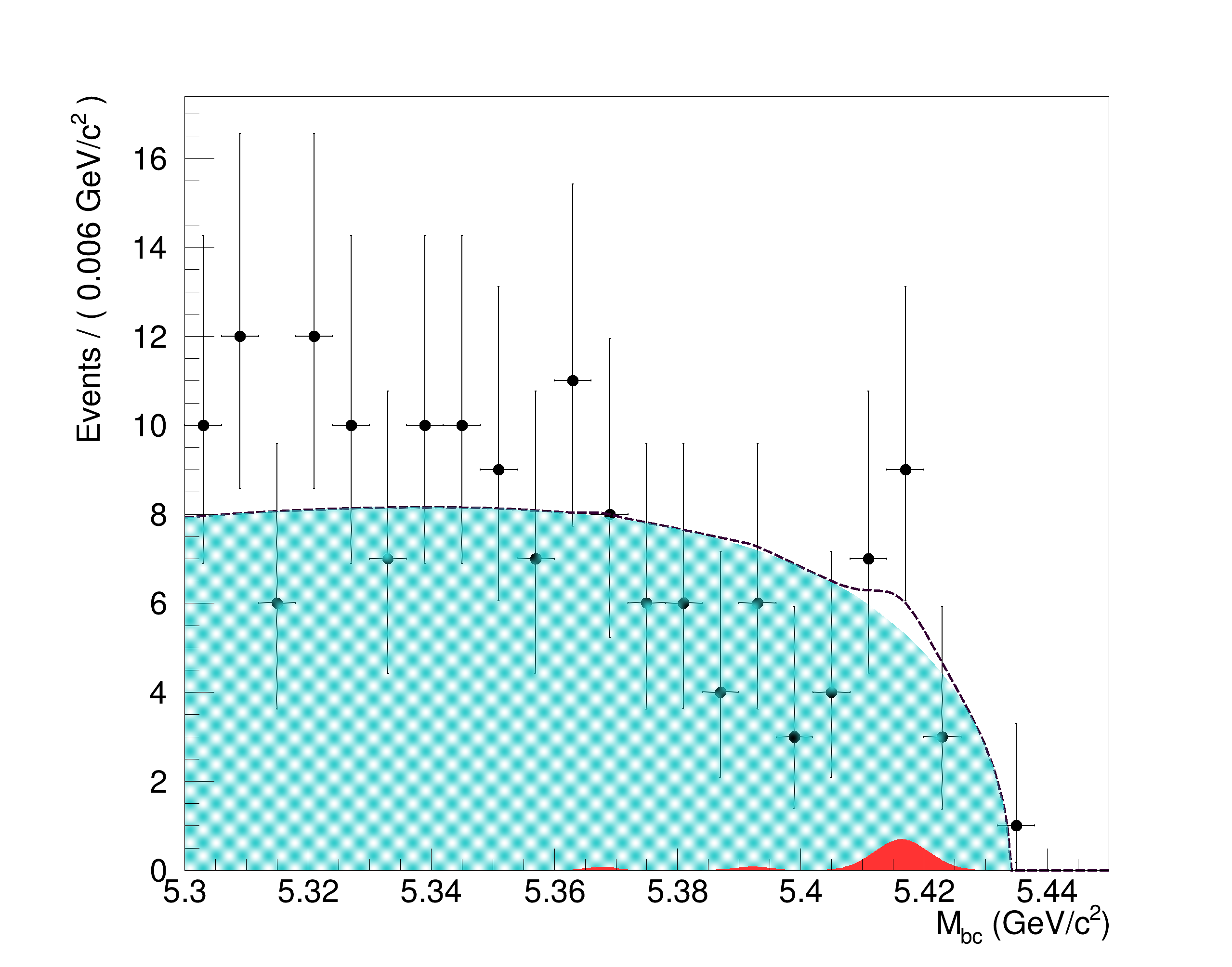}
	\caption{Sum of the fits to all $M(X_{s\bar{s}})$ bins overlaid on the $M_{\mathrm{bc}}$ distribution, for the decay $B_{s}^{0} \rightarrow \eta^{\prime} (\rightarrow \eta \pi^{+} \pi^{-}) X_{s\bar{s}}$ for $B_{s}^{0} \rightarrow \eta^{\prime} K^{+} K^{-} + n\pi$ submodes and $M(X_{s\bar{s}})$ $\leq$ 2.4 GeV/$c^{2}$ and with all selections applied.  The light blue shaded region is the sum of the background fits, the red shaded region is the sum of the signal fits, and the black dashed curve is the sum of the two.}
	\label{fig:r1}
\end{figure}

\begin{figure}[!htb]
		\includegraphics[scale=0.1]{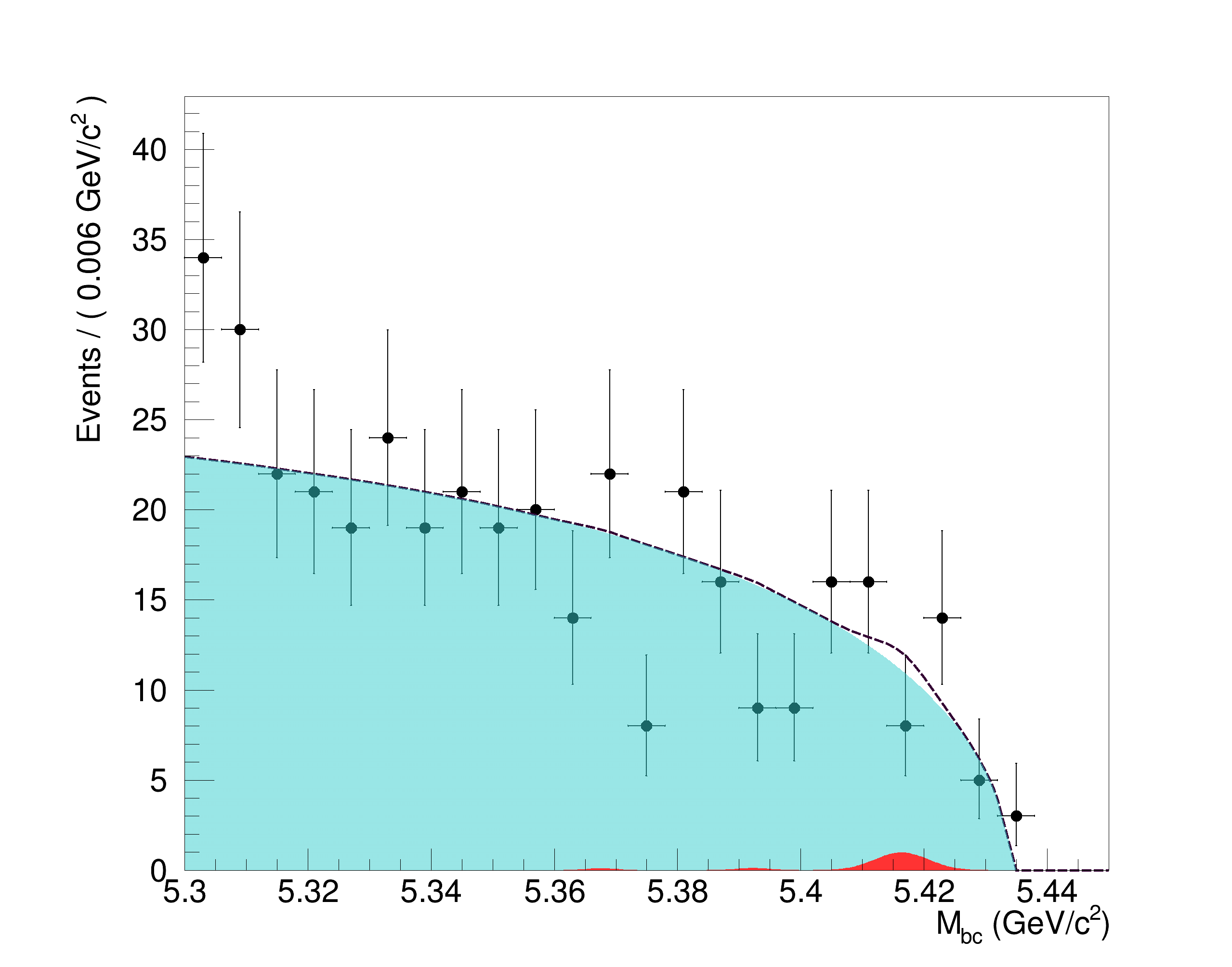}
	\caption{Sum of the fits to all $M(X_{s\bar{s}})$ bins overlaid on the $M_{\mathrm{bc}}$ distribution, for the decay $B_{s}^{0} \rightarrow \eta^{\prime} (\rightarrow \eta \pi^{+} \pi^{-}) X_{s\bar{s}}$ for $B_{s}^{0} \rightarrow \eta^{\prime} K^{\pm} K_{S}^{0} + n\pi$ submodes and $M(X_{s\bar{s}})$ $\leq$ 2.4 GeV/$c^{2}$ and with all selections applied.  The light blue shaded region is the sum of the background fits, the red shaded region is the sum of the signal fits, and the black dashed curve is the sum of the two.}
	\label{fig:r2}
\end{figure}

Figures \ref{fig:r1} and \ref{fig:r2} show the sum of the fits, whose results are listed in Tables \ref{table:1} and \ref{table:2}, respectively, overlaid on the data.  The central value for $\mathcal{B}$($B_{s}^{0} \rightarrow \eta^{\prime} X_{s\bar{s}}$) is estimated to be the weighted average of the total BF central values for $B_{s}^{0} \rightarrow \eta^{\prime} K^{+} K^{-} + n\pi$ and $B_{s}^{0} \rightarrow \eta^{\prime} K^{\pm} K^{0} + n\pi$.  These are obtained by summing the BFs listed in Tables \ref{table:1} and \ref{table:2}, for $B_{s}^{0} \rightarrow \eta^{\prime} K^{+} K^{-} + n\pi$ and $B_{s}^{0} \rightarrow \eta^{\prime} K^{\pm} K^{0} + n\pi$, respectively.  The weights for the average central value are obtained from the statistical uncertainties.  

The dominant uncertainties are due to the $X_{s\bar{s}}$ fragmentation.  Other systematic uncertainties include neural network selection, uncertainties related to track finding and identification, best candidate selection, neutral meson reconstruction, subdecay branching fractions, $\Upsilon(5S)$ production models, and the number of $B_{s}^{0}$$\bar{B}_{s}^{0}$ pairs.  A detailed discussion of the uncertainties is given in the accompanying appendix.  Systematic uncertainties are added in quadrature; fragmentation model (FM) \footnote{The term ``FM'' is used in this paper but is known as ``model'' in Ref. \cite{PhysRevLett.93.061801}.} uncertainties are added linearly within a class and for the final weighted average, these class sums are added in quadrature.

The statistical significance in each $X_{s\bar{s}}$ mass bin is calculated as $\mathcal{S} = \sqrt{-2\ln(\mathcal{L}_{0}/\mathcal{L}_{\rm max})}$, where $\mathcal{L}_{0}$ is the likelihood at zero signal yield and $\mathcal{L}_{\rm max}$ is the maximum likelihood.  No statistically significant excess of events is observed in any $X_{s\bar{s}}$ mass bin.  We set an upper limit on the partial BF (a BF with the requirement $M(X_{s\bar{s}})$ $\leq$ 2.4 GeV/$c^{2}$) at 90\% confidence level by integrating a Gaussian likelihood function whose standard deviation is estimated by the sum in quadrature of the positive statistical and systematic uncertainties.    The standard deviation, $\sigma$, is approximately 8.6 $\times$ 10$^{-4}$.  The integral is restricted to the physically allowed region above zero, giving an upper limit on $\mathcal{B}$($B_{s}^{0} \rightarrow \eta^{\prime} X_{s\bar{s}}$).  As a result, 1.68$\sigma$ is added to the weighted average central value to obtain the 90\% confidence level upper limit.

\begin{table}
	\begin{center}
		\caption{Results for the $B_{s}^{0} \rightarrow \eta^{\prime} K^{+} K^{-} + n\pi$ submodes, from the 121.4 fb$^{-1}$ $\Upsilon(5S)$ data set; the table contains the $M(X_{s\bar{s}})$ bin in units of GeV/$c^2$, corrected reconstruction efficiency ($\epsilon^{\prime}$), number of fitted signal events $N_{\rm sig}$, and $\mathcal{B}$, the central value of the partial BF.
}
\scalebox{0.95}{
		\begin{tabular}{ c  c  c  c }
	\hline
	\hline
	$M(X_{s\bar{s}})$ & $\epsilon^{\prime}$(\%) & $N_{\rm sig}$ & $\mathcal{B}(B_{s}^{0} \rightarrow \eta^{\prime} X_{s\bar{s}})$ (10$^{-4}$) \\ \hline
	1.0 - 1.2 & 3.60 $\pm$ 0.08 & 0.4$_{-1.9}^{+2.6}$ & 0.05$^{+0.30}_{-0.22}$ (stat.) $^{+0.004}_{-0.005}$ (syst.) \\ %\hline
	1.2 - 1.4 & 2.82 $\pm$ 0.08 & 0.08$_{-1.7}^{+2.4}$ & 0.01$^{+0.36}_{-0.28}$ (stat.) $^{+0.001}_{-0.001}$ (syst.) \\ %\hline
	1.4 - 1.6 & 0.90 $\pm$ 0.04 & 0.7$_{-1.8}^{+2.5}$ & 0.3$^{+1.1}_{-0.8}$ (stat.) $^{+0.04}_{-0.04}$ (syst.) \\ %\hline
	1.6 - 1.8 & 0.54 $\pm$ 0.03 & 0.4$_{-1.4}^{+2.1}$  & 0.3$^{+1.6}_{-1.1}$ (stat.) $^{+0.05}_{-0.1}$ syst.) \\ %\hline
	1.8 - 2.0 & 0.34 $\pm$ 0.03 & 1.4$_{-2.0}^{+2.6}$ & 1.7$^{+3.3}_{-2.5}$ (stat.) $^{+0.4}_{-0.6}$ (syst.)  \\ %\hline
	2.0 - 2.2 & 0.22 $\pm$ 0.02 & 0.3$_{-3.4}^{+3.7}$ & 0.6$^{+7.1}_{-6.4}$ (stat.) $^{+0.2}_{-0.2}$ (syst.) \\ %\hline
	2.2 - 2.4 & 0.14 $\pm$ 0.02 & $-$2.3$_{-3.4}^{+3.8}$ & $-$7.0$^{+11.6}_{-10.4}$ (stat.) $^{+1.7}_{-4.1}$ (syst.)  \\ \hline \hline
\end{tabular}
}
		\label{table:1}
	\end{center}
\end{table}

\begin{table}
	\begin{center}
		\caption{Results for the $B_{s}^{0} \rightarrow \eta^{\prime} K^{\pm} K^{0}_{S} + n\pi$ submodes, from the 121.4 fb$^{-1}$ $\Upsilon(5S)$ data set; rows with dashes indicate bins where no events, background or signal, were found; the table contains the $M(X_{s\bar{s}})$ bin in units of GeV/$c^2$, corrected reconstruction efficiency ($\epsilon^{\prime}$), number of fitted signal events $N_{\rm sig}$, and $\mathcal{B}$, the central value of the partial BF.}
\scalebox{0.95}{
	\begin{tabular}{ c  c  c  c }
	\hline
	\hline
	$M(X_{s\bar{s}})$ & $\epsilon^{\prime}$(\%) & $N_{\rm sig}$ & $\mathcal{B}(B_{s}^{0} \rightarrow \eta^{\prime} X_{s\bar{s}})$ (10$^{-4}$) \\ \hline
	1.0 - 1.2 & 0.016 $\pm$ 0.006 & 0.0 & - \\ %\hline
	1.2 - 1.4 & 0.24 $\pm$ 0.02 & 0.3$_{-0.8}^{+1.4}$ & 0.5$^{+2.5}_{-1.5}$ (stat.) $^{+0.1}_{-0.04}$ (syst.)  \\ %\hline
	1.4 - 1.6 & 0.86 $\pm$ 0.04 & 2.0$_{-2.2}^{+3.0}$ & 1.0$^{+1.4}_{-1.1}$ (stat.) $^{+0.1}_{-0.07}$ (syst.)  \\ %\hline
	1.6 - 1.8 & 0.65 $\pm$ 0.04 & 1.2$_{-2.6}^{+3.3}$ & 0.8$^{+2.1}_{-1.6}$ (stat.) $^{+0.1}_{-0.1}$ (syst.)  \\ %\hline
	1.8 - 2.0 & 0.45 $\pm$ 0.03 & 4.8$_{-3.4}^{+4.2}$ & 4.4$^{+3.9}_{-3.1}$ (stat.) $^{+0.9}_{-0.7}$ (syst.)  \\ %\hline
	2.0 - 2.2 & 0.36 $\pm$ 0.03 & $-$2.4$_{-3.2}^{+3.9}$ & $-$2.8$^{+4.6}_{-3.8}$ (stat.) $^{+0.9}_{-0.7}$ (syst.)  \\ %\hline
	2.2 - 2.4 & 0.16 $\pm$ 0.02 & $-$1.1$_{-2.9}^{+3.6}$ & $-$2.6$^{+8.9}_{-7.1}$ (stat.) $^{+0.2}_{-1.9}$ (syst.) \\ \hline \hline
	\end{tabular}
}
		\label{table:2}
	\end{center}
\end{table}

The central value of the BF is $\mathcal{B}$($B_{s}^{0} \rightarrow \eta^{\prime} X_{s\bar{s}}$) = [$-$0.7 $\pm$ 8.1 (stat.) $\pm$ 0.7 (syst.) $_{-6.0}^{+3.0}$ (FM) $\pm$ 0.1 (N$_{B^{0(*)}_{s}\bar{B}^{0(*)}_{s}}$)] $\times$ 10$^{-4}$ for $M(X_{s\bar{s}})$ $\leq$ 2.4 GeV/$c^{2}$.
The FM uncertainty is obtained by considering alternate sets of $X_{s\bar{s}}$ fragmentation parameter values in $\textsc{Pythia}$ and redetermining the signal reconstruction efficiency \footnote{Estimates of the FM uncertainty are given in the appendix.}.

The corresponding upper limit at 90\% confidence level on the partial BF, including all uncertainties, is 1.4 $\times$ 10$^{-3}$ for $M(X_{s\bar{s}})$ $\leq$ 2.4 GeV/$c^{2}$.  If $SU(3)$ symmetry holds, then the BFs of $B \rightarrow \eta^{\prime} X_{s}$ and $B_{s}^{0} \rightarrow \eta^{\prime} X_{s\bar{s}}$ would be equivalent and their ratio, $\mathcal{R}$($\eta^{\prime}$) = $\mathcal{B}$($B_{s}^{0} \rightarrow \eta^{\prime} X_{s\bar{s}}$)/$\mathcal{B}$($B \rightarrow \eta^{\prime} X_{s}$) would be close to 1 \cite{DATTA2020}.  The measured BF for the decay $B \rightarrow \eta^{\prime} X_{s}$ is [3.9 $\pm$ 0.8 (stat.) $\pm$ 0.5 (syst.) $\pm$ 0.8 (model)] $\times$ 10$^{-4}$ \cite{PhysRevLett.93.061801}.  Using this and the weighted average BF given previously for $B_{s}^{0} \rightarrow \eta^{\prime} X_{s\bar{s}}$, $\mathcal{R}$($\eta^{\prime}$) is approximately $-$0.2 $\pm 2.1 (\mathrm{stat.}) \pm 0.2 (\mathrm{syst.}) ^{+0.8}_{-1.5} (\mathrm{FM}) \pm 0.03 (N_{B^{0(*)}_{s}\bar{B}^{0(*)}_{s}})$.  Applying the same method as used to calculate the upper limit on $\mathcal{B}$($B_{s}^{0} \rightarrow \eta^{\prime} X_{s\bar{s}}$), the 90\% confidence level upper limit on $\mathcal{R}$($\eta^{\prime}$) is 3.5.

\begin{figure}[!htb]
	\includegraphics[scale=0.1]{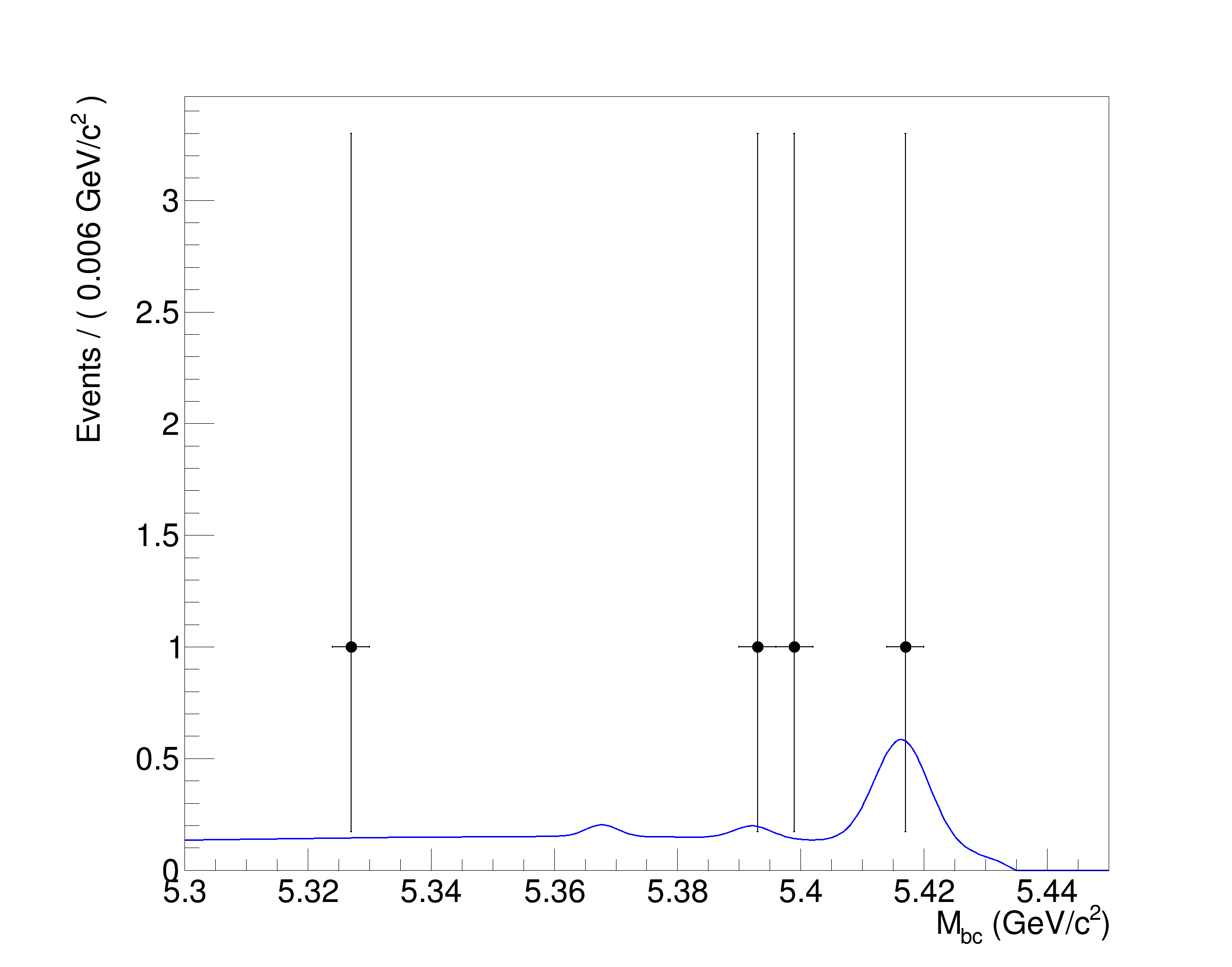}
	\caption{ $B_{s}^{0} \rightarrow \phi(\rightarrow K^{+} K^{-}) \eta^{\prime}$ decay results for M($X_{s\bar{s}}$) $\in$ $\pm$3$\sigma$ $\phi$ mass range}
	\label{fig:r3}
\end{figure}

As a by-product of the preceding measurement, we searched for the decay $B_{s}^{0} \rightarrow  \eta^{\prime} \phi$, with $\phi \rightarrow K^{+} K^{-}$.  This decay was searched for in the $X_{s\bar{s}}$ mass subrange $M(X_{s\bar{s}})$ $\in$ [1.006, 1.03] GeV/$c^{2}$ ($\pm$3$\sigma$ window).  From MC simulations, the reconstruction efficiency is determined to be 7.90 $\pm$ 0.03\%.  No statistically significant signal is found and the upper limit at 90\% confidence level is determined to be 3.6$\times$10$^{-5}$.  The result from fitting is shown in Fig. \ref{fig:r3}.  LHCb determines the upper limit at 90\% confidence level to be 8.2$\times$10$^{-7}$ \cite{Aaij:2016yuv}.

To conclude, we set an upper limit on the partial BF for the decay $B_{s}^{0} \rightarrow \eta^{\prime} X_{s\bar{s}}$, for $M(X_{s\bar{s}})$ $\leq$ 2.4 GeV/$c^{2}$.  Including all uncertainties, the upper limit at 90\% confidence level is determined to be 1.4 $\times$ 10$^{-3}$.  This is the first result for the inclusive decay $B_{s}^{0} \rightarrow \eta^{\prime} X_{s\bar{s}}$ and should motivate further studies, both experimental and theoretical, of inclusive $B_{s}^{0}$ meson processes and $SU(3)$ symmetries.

\

\begin{acknowledgments}
%----------- Long version, for most papers ----------- 
We thank the KEKB group for the excellent operation of the
accelerator; the KEK cryogenics group for the efficient
operation of the solenoid; and the KEK computer group, and the Pacific Northwest National
Laboratory (PNNL) Environmental Molecular Sciences Laboratory (EMSL)
computing group for strong computing support; and the National
Institute of Informatics, and Science Information NETwork 5 (SINET5) for
valuable network support.  We acknowledge support from
the Ministry of Education, Culture, Sports, Science, and
Technology (MEXT) of Japan, the Japan Society for the 
Promotion of Science (JSPS), and the Tau-Lepton Physics 
Research Center of Nagoya University; 
the Australian Research Council including grants
DP180102629, % Sevior
DP170102389, % Varvell
DP170102204, % Yabsley
DP150103061, % Urquijo
FT130100303; % Urquijo;
Austrian Federal Ministry of Education, Science and Research (FWF) and
FWF Austrian Science Fund No.~P~31361-N36;
the National Natural Science Foundation of China under Contracts
No.~11435013,  %Zhen-An Liu
No.~11475187,  %Chang-Zheng Yuan
No.~11521505,  %Chang-Zheng Yuan
No.~11575017,  %Cheng-Ping Shen
No.~11675166,  %Wen-Biao Yan
No.~11705209;  %Yi-Ming Li
Key Research Program of Frontier Sciences, Chinese Academy of Sciences (CAS), Grant No.~QYZDJ-SSW-SLH011; % Chang-Zheng Yuan
the  CAS Center for Excellence in Particle Physics (CCEPP); %Chang-Zheng Yuan,
the Shanghai Pujiang Program under Grant No.~18PJ1401000;  %Tao Luo
the Shanghai Science and Technology Committee (STCSM) under Grant No.~19ZR1403000; %Xiaolong Wang
the Ministry of Education, Youth and Sports of the Czech
Republic under Contract No.~LTT17020;
Horizon 2020 ERC Advanced Grant No.~884719 and ERC Starting Grant No.~947006 ``InterLeptons'' (European Union);
the Carl Zeiss Foundation, the Deutsche Forschungsgemeinschaft, the
Excellence Cluster Universe, and the VolkswagenStiftung;
the Department of Atomic Energy (Project Identification No. RTI 4002) and the Department of Science and Technology of India; 
the Istituto Nazionale di Fisica Nucleare of Italy; 
National Research Foundation (NRF) of Korea Grant
Nos.~2016R1\-D1A1B\-01010135, 2016R1\-D1A1B\-02012900, 2018R1\-A2B\-3003643,
2018R1\-A6A1A\-06024970, 2018R1\-D1A1B\-07047294, 2019K1\-A3A7A\-09033840,
2019R1\-I1A3A\-01058933;
Radiation Science Research Institute, Foreign Large-size Research Facility Application Supporting project, the Global Science Experimental Data Hub Center of the Korea Institute of Science and Technology Information and KREONET/GLORIAD;
the Polish Ministry of Science and Higher Education and 
the National Science Center;
the Ministry of Science and Higher Education of the Russian Federation, Agreement 14.W03.31.0026, % from 15.02.2018
and the HSE University Basic Research Program, Moscow; % from 15.04.2021
University of Tabuk research grants
S-1440-0321, S-0256-1438, and S-0280-1439 (Saudi Arabia);
the Slovenian Research Agency Grant Nos. J1-9124 and P1-0135;
Ikerbasque, Basque Foundation for Science, Spain;
the Swiss National Science Foundation; 
the Ministry of Education and the Ministry of Science and Technology of Taiwan;
and the United States Department of Energy and the National Science Foundation.

\end{acknowledgments}

\clearpage

\appendix*
\section{DISCUSSION OF SYSTEMATIC UNCERTAINTIES}

The upper limits at 90\% confidence level up to a given $X_{s\bar{s}}$ mass bin are given in Table \ref{table:rescumulative}.

\begin{table}[h]
	\begin{center}
		\caption{$\mathcal{B}_{UL}^{90\%} \leq$ $M(X_{s\bar{s}})$ 90\% upper limits.  Upper limit per bin corresponds to the upper limit up to and including that bin in units of $M(X_{s\bar{s}})$.}
		\begin{tabular}{ c c c }
			\hline
			\hline
			\textbf{$M(X_{s\bar{s}})$} & $\mathcal{B}(B_{s}^{0} \rightarrow \eta^{\prime} X_{s\bar{s}})$ (10$^{-4}$) & $\mathcal{B}_{UL}^{90\%}$ (10$^{-4}$) \\ \hline
			1.2 & 0.05 $\pm$ 0.26 (stat.) $^{+0.01}_{-0.01}$ (syst.) & 0.4  \\ %\hline
			1.4 & 0.08 $\pm$ 0.40 (stat.) $^{+0.10}_{-0.04}$ (syst.) & 0.7  \\ %\hline
			1.6 & 0.6 $\pm$ 1.0 (stat.) $^{+0.2}_{-0.1}$ (syst.) & 1.9  \\ %\hline
			1.8 & 1.1 $\pm$ 1.5 (stat.) $^{+0.3}_{-0.3}$ (syst.) & 3.1  \\ %\hline
			2.0 & 3.8 $\pm$ 2.7 (stat.) $^{+1.4}_{-1.3}$ (syst.) & 7.6  \\ %\hline
			2.2 & 3.4 $\pm$ 4.8 (stat.) $^{+2.2}_{-1.8}$ (syst.) & 11.1  \\ %\hline
			2.4 & $-$0.7 $\pm$ 8.1 (stat.) $^{+3.1}_{-6.0}$ (syst.) & 13.8  \\ \hline \hline
		\end{tabular}
		\label{table:rescumulative}
	\end{center}
\end{table}

Additive systematic uncertainties are from the PDF parameterization and fit bias.  The parameters of the Gaussian signal PDF are allowed to float within their 1$\sigma$ errors (determined from the $B_{s}^{0} \rightarrow D_{s}^{-} \rho^{+}$ control fit to the $\Upsilon(5S)$ data) and the $\Upsilon(5S)$ data are refitted for the signal yield.  The difference in signal yield between the fixed and floated parameterization is taken as the PDF uncertainty.  The same is done for the background ARGUS PDF.

The fit bias uncertainty is determined by generating and fitting 5000 MC pseudoexperiments for several assumptions of the branching fraction.  This is done using RooStats \cite{moneta2010roostats}.  The number of fitted signal events versus the number of generated signal events is fitted with a first-order polynomial and the offset from zero of the fit along the y-axis is taken as the uncertainty due to fit bias.  The fit bias uncertainty is less than one event.  The PDF and fit bias uncertainties are added in quadrature for a total additive systematic uncertainty.  This is combined with the statistical errors and quoted as the first uncertainty in Tables I and II in the main report.  For $B_{s}^{0} \rightarrow \eta^{\prime} K^{\pm} K^{0}_{S} + n\pi$, an uncertainty of 1.1 (26\%  of the fitted, positive statistical uncertainty) and 1.3 (34\%) events are obtained in $X_{s\bar{s}}$ mass bins 1.8 - 2.0 GeV/$c^{2}$ and 2.0 - 2.2 GeV/$c^{2}$, respectively.  All others had uncertainties of less than one event.  For $B_{s}^{0} \rightarrow \eta^{\prime} K^{+} K^{-} + n\pi$, the 1.6 - 1.8 GeV/$c^{2}$, 1.8 - 2.0 GeV/$c^{2}$, 2.0 - 2.2 GeV/$c^{2}$, and 2.2 - 2.4 GeV/$c^{2}$ bins have uncertainties of 1.0 (55\%), 1.2 (54\%), 3.1 (156\%), and 3.0 (132\%) events, respectively.   All other mass bins each have an uncertainty of less than one event.  Additive systematic uncertainties are added in quadrature with the asymmetric fit errors on the signal yield.

Multiplicative systematic uncertainties due to the fragmentation model (FM) of $X_{s\bar{s}}$ by $\textsc{Pythia}$ 6 \cite{Sj_strand_2006} are obtained by varying a group of $\textsc{Jetset}$ parameters - PARJ(1, 2, 3, 4, 11, 12, 13, 25, 26), described in Table \ref{table:param_meaning} - which are varied together away from the standard Belle default to reduce and enhance the (uncorrected) reconstruction efficiency, giving two sets of parameters for each $X_{s\bar{s}}$ bin.   These alternative tunings (``AT") are given in Table \ref{table:pythia_syst}.  They are motivated by the parameter studies in other inclusive $B$ analyses \cite{Buckley_2009, bn1066,Nishimura:2009ae,PhysRevD.86.052012,PhysRevD.91.052004}.  The uncertainty is determined from the fractional change in efficiency with respect to the Belle default parameters.  This procedure includes the effect of the change in the proportion of unreconstructed modes.  If no increase or decrease in efficiency is found then an uncertainty of zero is assigned.  Values for the FM uncertainty, in each $X_{s\bar{s}}$ mass bin, are given in Tables \ref{table:fmsystsummary} and \ref{table:fmsystsummary2}, obtained from the (uncorrected) efficiencies in Tables \ref{table:pythia_effs_nonks} and \ref{table:pythia_effs_ks}.

\begin{table}
	\begin{center}
		\caption{$\textsc{Jetset}$ parameter descriptions}
		\scalebox{0.90}{
			\begin{tabular}{ c c }
				\hline
				\hline
				Parameter & Description \\ \hline
				PARJ(1) & baryon suppression  \\ %\hline
				PARJ(2) &  $s$ vs $u,d$ quark suppression \\ %\hline
				PARJ(3) & $s$ quark further suppression \\ %\hline
				PARJ(4) & spin-1 diquark suppression vs spin-0 diquarks \\ %\hline
				PARJ(11) & probability of spin-1 light mesons \\ %\hline
				PARJ(12) & probability of spin-1 strange meson \\ %\hline
				PARJ(13) & probability of spin-1 meson with $c$ or heavier quark \\ %\hline
				PARJ(25) & $\eta$ suppression factor \\ %\hline
				PARJ(26) & $\eta^{\prime}$ suppression factor \\ \hline \hline
			\end{tabular}
		}
		\label{table:param_meaning}
	\end{center}
\end{table}

\begin{table}
	\begin{center}
		\caption{$\textsc{Jetset}$ parameters used to tune the fragmentation of the $X_{{s}\bar{s}}$ system in $\textsc{Pythia}$.  Alternative tunings (AT) AT1 and AT2 are used to obtain the systematic uncertainties due to fragmentation.}
		\scalebox{0.90}{
			\begin{tabular}{ c  c  c  c  c  c }
				\hline
				\hline
				Parameter & Standard & Ref. \cite{Buckley_2009} & Ref. \cite{bn1066} & AT1 & AT2 \\ \hline
				PARJ(1)  & 0.1 & 0.073 & 0.073 & 0.2 & 0.1 \\ %\hline
				PARJ(2)  & 0.3 & 0.2 & 1 & 0.2 & 0.4 \\ %\hline
				PARJ(3)  & 0.4 & 0.94 & 0.94 & 0.4 & 0.4 \\ %\hline
				PARJ(4)  & 0.05 & 0.032 & 0.032 & 0.264 & 0.008 \\ %\hline
				PARJ(11)  & 0.5 & 0.31 & 0.01 & 0.9 & 0.1 \\ %\hline
				PARJ(12)  & 0.6 & 0.4 & 0.01 & 0.6 & 0.6 \\ %\hline
				PARJ(13)  & 0.75 & 0.54 & 0.54 & 0.75 & 0.75 \\ %\hline
				PARJ(25)  & 1 & 0.63 & 1 & 0.1 & 1 \\ %\hline
				PARJ(26)  & 0.4 & 0.12 & 0.12 & 0.4 & 0.12 \\ \hline \hline
			\end{tabular}
		}
		\label{table:pythia_syst}
	\end{center}
\end{table}

\begin{table}
	\begin{center}
		\caption{Comparison of uncorrected reconstruction efficiencies and their associated relative systematic uncertainties (\%) between $\textsc{Pythia}$ tunings (Standard, AT1, and AT2) given in Table \ref{table:pythia_syst}, used in systematic uncertainty estimation; tuning is done in 0.2 GeV/$c^{2}$ $X_{s\bar{s}}$ mass bins for $B^{0}_{s} \rightarrow \eta^{\prime}  K^{+}K^{-} + n\pi$ modes.}
		\scalebox{0.95}{
			\begin{tabular}{ c c c c }
				\hline
				\hline
				$M(X_{s\bar{s}})$ & Standard & AT1 & AT2  \\ \hline
				%0.8-1.0 & 5.6 $\pm$ 0.41 & 4.9 $\pm$ 0.4 & 5.0 $\pm$ 0.4   \\ \hline
				1.0-1.2 & 3.76 $\pm$ 0.09 & 3.99 $\pm$ 0.09 & 3.75 $\pm$ 0.09  \\ %\hline
				1.2-1.4 & 2.96 $\pm$ 0.08 & 3.04 $\pm$ 0.08 & 2.77 $\pm$ 0.08  \\ %\hline
				1.4-1.6 & 0.96 $\pm$ 0.05 & 1.04 $\pm$ 0.05 & 0.89 $\pm$ 0.04   \\ %\hline
				1.6-1.8 & 0.58 $\pm$ 0.04 & 0.78 $\pm$ 0.04 & 0.49 $\pm$ 0.03   \\ %\hline
				1.8-2.0 & 0.36 $\pm$ 0.03 & 0.48 $\pm$ 0.03 & 0.29 $\pm$ 0.03   \\ %\hline
				2.0-2.2 & 0.24 $\pm$ 0.02 & 0.32 $\pm$ 0.03 & 0.17 $\pm$ 0.02   \\ %\hline
				2.2-2.4 & 0.15 $\pm$ 0.02 & 0.23 $\pm$ 0.02 & 0.11 $\pm$ 0.02   \\ \hline \hline
			\end{tabular}
		}
		\label{table:pythia_effs_nonks}
	\end{center}
\end{table}

\begin{table}
	\begin{center}
		\caption{Comparison of uncorrected reconstruction efficiencies and their associated relative systematic uncertainties (\%) between $\textsc{Pythia}$ tunings (Standard, AT1, and AT2) given in Table \ref{table:pythia_syst}, used in systematic uncertainty estimation; tuning is done in 0.2 GeV/$c^{2}$ $X_{s\bar{s}}$ mass bins for $B^{0}_{s} \rightarrow \eta^{\prime}  K^{\pm}K_{S}^{0} + n\pi$ modes.}
		\scalebox{0.95}{
			\begin{tabular}{ c c c c }
				\hline
				\hline
				$M(X_{s\bar{s}})$ & Standard & AT1 & AT2 \\ \hline
				%0.8-1.0 & 0.0 & 0.0 & 0.0 \\ \hline
				1.0-1.2 & 0.016 $\pm$ 0.006 & 0.001 $\pm$ 0.004 & 0.012 $\pm$ 0.006 \\ %\hline
				1.2-1.4 & 0.25 $\pm$ 0.02 & 0.26 $\pm$ 0.03 & 0.21 $\pm$ 0.02  \\ %\hline
				1.4-1.6 & 0.90 $\pm$ 0.05 & 0.79 $\pm$ 0.04 & 0.84 $\pm$ 0.05  \\ %\hline
				1.6-1.8 & 0.68 $\pm$ 0.04 & 0.76 $\pm$ 0.04 & 0.60 $\pm$ 0.04  \\ %\hline
				1.8-2.0 & 0.48 $\pm$ 0.04 & 0.55 $\pm$ 0.04 & 0.38 $\pm$ 0.03  \\ %\hline
				2.0-2.2 & 0.38 $\pm$ 0.03 & 0.47 $\pm$ 0.04 & 0.26 $\pm$ 0.03  \\ %\hline
				2.2-2.4 & 0.18 $\pm$ 0.03 & 0.32 $\pm$ 0.03 & 0.19 $\pm$ 0.03  \\ \hline \hline
			\end{tabular}
		}
		\label{table:pythia_effs_ks}
	\end{center}
\end{table}

\begin{table}[h]
	\begin{center}
		\caption{Summary of FM multiplicative systematic uncertainties for $B_{s}^{0} \rightarrow \eta^{\prime} K^{+} K^{-} + n\pi$.}
		\begin{tabular}{ c c }
			\hline
			\hline
			$M(X_{s\bar{s}})$ & FM (\%)  \\ \hline
			1.0 - 1.2 & $^{+0.4}_{-5.9}$ \\ %\hline
			1.2 - 1.4 & $^{+6.4}_{-2.8}$  \\ %\hline
			1.4 - 1.6 & $^{+8.0}_{-8.3}$  \\ %\hline
			1.6 - 1.8 & $^{+14.7}_{-35.3}$  \\ %\hline
			1.8 - 2.0 & $^{+21.1}_{-33.6}$    \\ %\hline
			2.0 - 2.2 & $^{+28.7}_{-37.4}$  \\ %\hline
			2.2 - 2.4 & $^{+23.7}_{-58.2}$  \\ \hline \hline
		\end{tabular}
		\label{table:fmsystsummary}
	\end{center}
\end{table}

\begin{table}[h]
	\begin{center}
		\caption{Summary of FM multiplicative systematic uncertainties for $B_{s}^{0} \rightarrow \eta^{\prime} K^{\pm} K^{0}_{S} + n\pi$.}
		\begin{tabular}{ c c }
			\hline
			\hline
			$M(X_{s\bar{s}})$ & FM (\%)  \\ \hline
			1.0 - 1.2 & $^{+23.7}_{-0.0}$   \\ %\hline
			1.2 - 1.4 & $^{+18.3}_{-2.3}$   \\ %\hline
			1.4 - 1.6 & $^{+6.6}_{-0.0}$   \\ %\hline
			1.6 - 1.8 & $^{+12.5}_{-10.5}$  \\ %\hline
			1.8 - 2.0 & $^{+20.2}_{-14.4}$   \\ %\hline
			2.0 - 2.2 & $^{+30.7}_{-23.2}$  \\ %\hline
			2.2 - 2.4 & $^{+0.0}_{-74.5}$  \\ \hline \hline
		\end{tabular}
		\label{table:fmsystsummary2}
	\end{center}
\end{table}

From the signal MC that is generated and used to determine signal reconstruction efficiency, the proportion of unreconstructed modes is determined by searching in the generated signal MC for modes that contain an $X_{s\bar{s}}$ decay submode but fall outside the criteria for a reconstructed submode, i.e. submodes that contain more than one $\pi^{0}$, modes with a $K_{L}^{0}$, or modes with more than six daughter particles (excluding the $\eta^{\prime}$).  The proportion of unreconstructed events, defined as $N_{\mathrm{UR}}$/($N_{\mathrm{UR}}$ + $N_{R}$), where $N_{\mathrm{UR}}$ is the number of generated events from unreconstructed signal modes in signal MC, and $N_{R}$ is the number of generated events from reconstructed modes.  For $B_{s}^{0} \rightarrow \eta^{\prime} K^{+} K^{-} + n\pi$, 1.1\% of events are unreconstructed in the 1.4-1.6 GeV/$c^{2}$ bin, increasing monotonically to 14.5\% in the 2.2-2.4 GeV/$c^{2}$ bin.  For $B_{s}^{0} \rightarrow \eta^{\prime} K^{\pm} K^{0} + n\pi$ modes, as they are only reconstructed as $B_{s}^{0} \rightarrow \eta^{\prime} K^{\pm} K^{0}_{S} + n\pi$, there is a corresponding class of modes that involve a $K^{0}_{L}$ instead of a $K^{0}_{S}$.  This causes the proportion of generated signal events to be higher.  In the 1.0-1.2 GeV/$c^{2}$ bin, 48.1\% of reconstructable events are unreconstructed, due to unreconstructed $K^{0}_{L}$ modes.  This increases monotonically to 59.7\% in the 2.2-2.4 GeV/$c^{2}$ bin, of which 84\% is due to unreconstructed $K^{0}_{L}$ modes.  Using the same signal MC, it is also found that the signal cross-feed efficiency is less than 0.05\% in each $X_{s\bar{s}}$ mass bin and is included in the multiplicative systematic uncertainties.

The $B_{s}^{0} \rightarrow D_{s}^{-} \rho^{+}$ control sample is used to determine the systematic uncertainty with respect to the neural network (NN) selection.  This uncertainty is obtained by determining the signal yield with and without the neural network selection in both MC and data. The double ratio of these results is determined and its absolute difference from unity is used as the systematic uncertainty.  This gives an uncertainty of 6.5\% for $B_{s}^{0} \rightarrow \eta^{\prime} K^{+} K^{-} + n\pi$ and 2.1\% for $B_{s}^{0} \rightarrow \eta^{\prime} K^{\pm} K^{0}_{S} + n\pi$.  The control sample $B_{s}^{0} \rightarrow D_{s}\rho$ is also used to obtain the uncertainty for best candidate selection (BCS).  The uncertainty is obtained by determining the signal yield with and without best candidate selection in both MC and data.  The double ratio of these results is determined and its absolute difference from unity is used as the systematic uncertainty.  This gives an uncertainty of 1.0\% for $B_{s}^{0} \rightarrow \eta^{\prime} K^{+} K^{-} + n\pi$ and 4.4\% for $B_{s}^{0} \rightarrow \eta^{\prime} K^{\pm} K^{0}_{S} + n\pi$, using the neural network selection of these associated classes of signal modes.  The uncertainty for the reconstruction of $\eta \rightarrow \gamma \gamma$ and $\pi^{0} \rightarrow \gamma \gamma$ is 3.0\% \cite{PhysRevD.92.011101}.

The uncertainty on charged track reconstruction is 0.35\% per track \cite{PhysRevD.89.072009}.  The uncertainty on the efficiency to identify charged kaons and pions is a function of their momenta and polar angles.  The uncertainty for $K^{\pm}$ and $\pi^{\pm}$ identification is 0.95\% and 1.8\%, respectively.  The $K_{S}^{0}$ reconstruction uncertainty is 1.6\% \cite{PhysRevLett.119.171801}.  The total track uncertainty, for each source, per $X_{s\bar{s}}$ mass bin, is obtained by determining the average charged kaon and charged pion multiplicity ($M$) in signal MC and multiplying the uncertainty by that multiplicity, e.g. $M$(0.182).  These uncertainties are added linearly as they are uncertainties of common daughters of a single mother particle ($B_{s}^{0}$) and are thus correlated.

The $\Upsilon(5S)$ production model (PM) uncertainty leads to a fractional change in reconstruction efficiency of $B_{s}^{0*}\bar{B}_{s}^{0*}$ $S$-wave ($L = 0$) states in a $B \rightarrow D_{s} \pi$ control sample MC, with and without the model in \cite{abdesselam2016study2}, is implemented.  The uncertainty is approximately 0.2\% for $B_{s}^{0} \rightarrow \eta^{\prime} K^{\pm} K^{0} + n\pi$ and 1.1\% for $B_{s}^{0} \rightarrow \eta^{\prime} K^{\pm} K^{0}_{S} + n\pi$.  The uncertainty on the subdecay mode branching fractions $\mathcal{B}(\eta \rightarrow \gamma \gamma)$ and $\mathcal{B}(\eta^{\prime} \rightarrow \eta \pi \pi)$ are 0.2\% and 0.7\%, respectively \cite{Tanabashi:2018oca}.  Estimates of individual multiplicative systematic uncertainties are given in Table \ref{table:systsummary}.  Totals of these uncertainties are determined in individual $X_{s\bar{s}}$ mass bins. 

\begin{table}[h]
	\begin{center}
		\caption{Summary of multiplicative systematic uncertainties.  The uncertainties for particle identification and reconstruction are evaluated per $X_{s\bar{s}}$ mass bin.}
		\begin{tabular}{ c c }
			\hline
			\hline
			Uncertainty Source & Value (\%) \\ \hline
			$\pi^{0}$ reconstruction &   3.0  \\ %\hline 
			$K_{S}^{0}$ reconstruction &  1.6   \\ %\hline
			Charged track reconstruction &   0.4  \\ %\hline
			$K^{\pm}$ ID &  0.95   \\ %\hline
			$\pi^{\pm}$ ID &   1.3  \\ %\hline
			$\Upsilon(5S)$ PM ($B_{s}^{0} \rightarrow \eta^{\prime} K^{+} K^{-} + n\pi$) &   0.2  \\ %\hline
			$\Upsilon(5S)$ PM ($B_{s}^{0} \rightarrow \eta^{\prime} K^{\pm} K^{0}_{S} + n\pi$) &   1.1  \\ %\hline
			$\eta$ reconstruction &  3.0   \\ %\hline
			NN Selection ($B_{s}^{0} \rightarrow \eta^{\prime} K^{+} K^{-} + n\pi$) &  6.5   \\ %\hline
			NN Selection ($B_{s}^{0} \rightarrow \eta^{\prime} K^{\pm} K^{0}_{S} + n\pi$) &  2.1  \\ %\hline
			BCS ($B_{s}^{0} \rightarrow \eta^{\prime} K^{+} K^{-} + n\pi$) &   1.0  \\ %\hline
			BCS ($B_{s}^{0} \rightarrow \eta^{\prime} K^{\pm} K^{0}_{S} + n\pi$) &   4.4  \\ %\hline
			$\mathcal{B}(\eta \rightarrow \gamma \gamma)$ &  0.2   \\ %\hline
			$\mathcal{B}(\eta^{\prime} \rightarrow \eta \pi \pi)$ &  0.7   \\ %\hline
			$N_{B_{s}^{0(*)}\bar{B}_{s}^{0(*)}}$ &  18.3   \\ \hline \hline
		\end{tabular}
		\label{table:systsummary}
	\end{center}
\end{table}

%%%%%

%%%%%

\clearpage

\bibliography{bs2etapxss_post_proofs.bib}

\end{document}

%% file: pub574.tex
%%% Paper:    Bs -> eta' Xss
%%% Journal:  Physical Review Letters
%%% Contacts: S. Dubey (sdubey@hawaii.edu)
%%%           T. Browder (teb@phys.hawaii.edu)
%%% Non-responding authors or those who said NO are commented out.
%%% ====================================================================
%%% Click the RELOAD button on your web browser to see the updated file.
%%% ====================================================================
%%% Use \input{author} to insert this material into your latex file.
%%%%% Force institutions to appear in alphabetical order when typeset.
\noaffiliation
\affiliation{University of the Basque Country UPV/EHU, 48080 Bilbao}
%%%\affiliation{Beihang University, Beijing 100191}
\affiliation{University of Bonn, 53115 Bonn}
\affiliation{Brookhaven National Laboratory, Upton, New York 11973}
\affiliation{Budker Institute of Nuclear Physics SB RAS, Novosibirsk 630090}
\affiliation{Faculty of Mathematics and Physics, Charles University, 121 16 Prague}
%%%\affiliation{Chiba University, Chiba 263-8522}
\affiliation{Chonnam National University, Gwangju 61186}
\affiliation{University of Cincinnati, Cincinnati, Ohio 45221}
\affiliation{Deutsches Elektronen--Synchrotron, 22607 Hamburg}
%%%\affiliation{Duke University, Durham, North Carolina 27708}
\affiliation{University of Florida, Gainesville, Florida 32611}
%%%\affiliation{Department of Physics, Fu Jen Catholic University, Taipei 24205}
%%%\affiliation{Key Laboratory of Nuclear Physics and Ion-beam Application (MOE) and Institute of Modern Physics, Fudan University, Shanghai 200443}
\affiliation{Justus-Liebig-Universit\"at Gie\ss{}en, 35392 Gie\ss{}en}
%%%\affiliation{Gifu University, Gifu 501-1193}
%%%\affiliation{II. Physikalisches Institut, Georg-August-Universit\"at G\"ottingen, 37073 G\"ottingen}
\affiliation{SOKENDAI (The Graduate University for Advanced Studies), Hayama 240-0193}
\affiliation{Gyeongsang National University, Jinju 52828}
\affiliation{Department of Physics and Institute of Natural Sciences, Hanyang University, Seoul 04763}
\affiliation{University of Hawaii, Honolulu, Hawaii 96822}
\affiliation{High Energy Accelerator Research Organization (KEK), Tsukuba 305-0801}
\affiliation{J-PARC Branch, KEK Theory Center, High Energy Accelerator Research Organization (KEK), Tsukuba 305-0801}
\affiliation{Higher School of Economics (HSE), Moscow 101000}
\affiliation{Forschungszentrum J\"{u}lich, 52425 J\"{u}lich}
%%%\affiliation{Hiroshima Institute of Technology, Hiroshima 731-5193}
\affiliation{IKERBASQUE, Basque Foundation for Science, 48013 Bilbao}
%%%\affiliation{University of Illinois at Urbana-Champaign, Urbana, Illinois 61801}
%%%\affiliation{Indian Institute of Science Education and Research Mohali, SAS Nagar, 140306}
\affiliation{Indian Institute of Technology Bhubaneswar, Satya Nagar 751007}
\affiliation{Indian Institute of Technology Guwahati, Assam 781039}
\affiliation{Indian Institute of Technology Hyderabad, Telangana 502285}
\affiliation{Indian Institute of Technology Madras, Chennai 600036}
%%%\affiliation{Indiana University, Bloomington, Indiana 47408}
\affiliation{Institute of High Energy Physics, Chinese Academy of Sciences, Beijing 100049}
\affiliation{Institute of High Energy Physics, Vienna 1050}
\affiliation{Institute for High Energy Physics, Protvino 142281}
%%%\affiliation{Institute of Mathematical Sciences, Chennai 600113}
\affiliation{INFN - Sezione di Napoli, 80126 Napoli}
\affiliation{INFN - Sezione di Torino, 10125 Torino}
\affiliation{Advanced Science Research Center, Japan Atomic Energy Agency, Naka 319-1195}
\affiliation{J. Stefan Institute, 1000 Ljubljana}
%%%\affiliation{Kanagawa University, Yokohama 221-8686}
\affiliation{Institut f\"ur Experimentelle Teilchenphysik, Karlsruher Institut f\"ur Technologie, 76131 Karlsruhe}
\affiliation{Kavli Institute for the Physics and Mathematics of the Universe (WPI), University of Tokyo, Kashiwa 277-8583}
\affiliation{Kennesaw State University, Kennesaw, Georgia 30144}
%%%\affiliation{King Abdulaziz City for Science and Technology, Riyadh 11442}
\affiliation{Department of Physics, Faculty of Science, King Abdulaziz University, Jeddah 21589}
\affiliation{Kitasato University, Sagamihara 252-0373}
\affiliation{Korea Institute of Science and Technology Information, Daejeon 34141}
\affiliation{Korea University, Seoul 02841}
%%%\affiliation{Kyoto Sangyo University, Kyoto 603-8555}
%%%\affiliation{Kyoto University, Kyoto 606-8502}
\affiliation{Kyungpook National University, Daegu 41566}
\affiliation{Universit\'{e} Paris-Saclay, CNRS/IN2P3, IJCLab, 91405 Orsay}
%%%\affiliation{\'Ecole Polytechnique F\'ed\'erale de Lausanne (EPFL), Lausanne 1015}
\affiliation{P.N. Lebedev Physical Institute of the Russian Academy of Sciences, Moscow 119991}
\affiliation{Liaoning Normal University, Dalian 116029}
\affiliation{Faculty of Mathematics and Physics, University of Ljubljana, 1000 Ljubljana}
\affiliation{Ludwig Maximilians University, 80539 Munich}
\affiliation{Luther College, Decorah, Iowa 52101}
\affiliation{Malaviya National Institute of Technology Jaipur, Jaipur 302017}
%%%\affiliation{University of Malaya, 50603 Kuala Lumpur}
\affiliation{University of Maribor, 2000 Maribor}
\affiliation{Max-Planck-Institut f\"ur Physik, 80805 M\"unchen}
\affiliation{School of Physics, University of Melbourne, Victoria 3010}
\affiliation{University of Mississippi, University, Mississippi 38677}
\affiliation{University of Miyazaki, Miyazaki 889-2192}
%%%\affiliation{Moscow Physical Engineering Institute, Moscow 115409}
\affiliation{Graduate School of Science, Nagoya University, Nagoya 464-8602}
%%%\affiliation{Kobayashi-Maskawa Institute, Nagoya University, Nagoya 464-8602}
\affiliation{Universit\`{a} di Napoli Federico II, 80126 Napoli}
%%%\affiliation{Nara University of Education, Nara 630-8528}
\affiliation{Nara Women's University, Nara 630-8506}
%%%\affiliation{National Central University, Chung-li 32054}
\affiliation{National United University, Miao Li 36003}
\affiliation{Department of Physics, National Taiwan University, Taipei 10617}
\affiliation{H. Niewodniczanski Institute of Nuclear Physics, Krakow 31-342}
\affiliation{Nippon Dental University, Niigata 951-8580}
\affiliation{Niigata University, Niigata 950-2181}
%%%\affiliation{University of Nova Gorica, 5000 Nova Gorica}
\affiliation{Novosibirsk State University, Novosibirsk 630090}
%%%\affiliation{Okinawa Institute of Science and Technology, Okinawa 904-0495}
\affiliation{Osaka City University, Osaka 558-8585}
%%%\affiliation{Osaka University, Osaka 565-0871}
\affiliation{Pacific Northwest National Laboratory, Richland, Washington 99352}
\affiliation{Panjab University, Chandigarh 160014}
\affiliation{Peking University, Beijing 100871}
\affiliation{University of Pittsburgh, Pittsburgh, Pennsylvania 15260}
\affiliation{Punjab Agricultural University, Ludhiana 141004}
%%%\affiliation{Research Center for Electron Photon Science, Tohoku University, Sendai 980-8578}
\affiliation{Research Center for Nuclear Physics, Osaka University, Osaka 567-0047}
\affiliation{Meson Science Laboratory, Cluster for Pioneering Research, RIKEN, Saitama 351-0198}
%%%\affiliation{Theoretical Research Division, Nishina Center, RIKEN, Saitama 351-0198}
%%%\affiliation{RIKEN BNL Research Center, Upton, New York 11973}
%%%\affiliation{Saga University, Saga 840-8502}
\affiliation{Department of Modern Physics and State Key Laboratory of Particle Detection and Electronics, University of Science and Technology of China, Hefei 230026}
\affiliation{Seoul National University, Seoul 08826}
\affiliation{Showa Pharmaceutical University, Tokyo 194-8543}
\affiliation{Soochow University, Suzhou 215006}
\affiliation{Soongsil University, Seoul 06978}
%%%\affiliation{University of South Carolina, Columbia, South Carolina 29208}
%%%\affiliation{Stefan Meyer Institute for Subatomic Physics, Vienna 1090}
\affiliation{Sungkyunkwan University, Suwon 16419}
\affiliation{School of Physics, University of Sydney, New South Wales 2006}
\affiliation{Department of Physics, Faculty of Science, University of Tabuk, Tabuk 71451}
\affiliation{Tata Institute of Fundamental Research, Mumbai 400005}
%%%\affiliation{Excellence Cluster Universe, Technische Universit\"at M\"unchen, 85748 Garching}
\affiliation{Department of Physics, Technische Universit\"at M\"unchen, 85748 Garching}
\affiliation{School of Physics and Astronomy, Tel Aviv University, Tel Aviv 69978}
\affiliation{Toho University, Funabashi 274-8510}
\affiliation{Department of Physics, Tohoku University, Sendai 980-8578}
\affiliation{Earthquake Research Institute, University of Tokyo, Tokyo 113-0032}
\affiliation{Department of Physics, University of Tokyo, Tokyo 113-0033}
\affiliation{Tokyo Institute of Technology, Tokyo 152-8550}
%%%\affiliation{Tokyo Metropolitan University, Tokyo 192-0397}
%%%\affiliation{Tokyo University of Agriculture and Technology, Tokyo 184-8588}
\affiliation{Utkal University, Bhubaneswar 751004}
\affiliation{Virginia Polytechnic Institute and State University, Blacksburg, Virginia 24061}
\affiliation{Wayne State University, Detroit, Michigan 48202}
\affiliation{Yamagata University, Yamagata 990-8560}
\affiliation{Yonsei University, Seoul 03722}
\author{S.~Dubey}\affiliation{University of Hawaii, Honolulu, Hawaii 96822}
\author{T.~E.~Browder}\affiliation{University of Hawaii, Honolulu, Hawaii 96822} % Hawaii
% \author{A.~Abdesselam}\affiliation{Department of Physics, Faculty of Science, University of Tabuk, Tabuk 71451} % Tabuk
% \author{I.~Adachi}\affiliation{High Energy Accelerator Research Organization (KEK), Tsukuba 305-0801}\affiliation{SOKENDAI (The Graduate University for Advanced Studies), Hayama 240-0193} % KEK
% \author{K.~Adamczyk}\affiliation{H. Niewodniczanski Institute of Nuclear Physics, Krakow 31-342} % Krakow
% \author{J.~K.~Ahn}\affiliation{Korea University, Seoul 02841} % Korea
  \author{H.~Aihara}\affiliation{Department of Physics, University of Tokyo, Tokyo 113-0033} % Tokyo
  \author{S.~Al~Said}\affiliation{Department of Physics, Faculty of Science, University of Tabuk, Tabuk 71451}\affiliation{Department of Physics, Faculty of Science, King Abdulaziz University, Jeddah 21589} % Tabuk
% \author{K.~Arinstein}\affiliation{Budker Institute of Nuclear Physics SB RAS, Novosibirsk 630090}\affiliation{Novosibirsk State University, Novosibirsk 630090} % BINP
% \author{Y.~Arita}\affiliation{Graduate School of Science, Nagoya University, Nagoya 464-8602} % Nagoya
  \author{D.~M.~Asner}\affiliation{Brookhaven National Laboratory, Upton, New York 11973} % BNL
% \author{H.~Atmacan}\affiliation{University of Cincinnati, Cincinnati, Ohio 45221} % Cincinnati
% \author{V.~Aulchenko}\affiliation{Budker Institute of Nuclear Physics SB RAS, Novosibirsk 630090}\affiliation{Novosibirsk State University, Novosibirsk 630090} % BINP
  \author{T.~Aushev}\affiliation{Higher School of Economics (HSE), Moscow 101000} % HSE
  \author{R.~Ayad}\affiliation{Department of Physics, Faculty of Science, University of Tabuk, Tabuk 71451} % Tabuk
% \author{T.~Aziz}\affiliation{Tata Institute of Fundamental Research, Mumbai 400005} % Tata
  \author{V.~Babu}\affiliation{Deutsches Elektronen--Synchrotron, 22607 Hamburg} % DESY
  \author{S.~Bahinipati}\affiliation{Indian Institute of Technology Bhubaneswar, Satya Nagar 751007} % IITB
% \author{A.~M.~Bakich}\affiliation{School of Physics, University of Sydney, New South Wales 2006} % Sydney
% \author{Y.~Ban}\affiliation{Peking University, Beijing 100871} % Peking
% \author{E.~Barberio}\affiliation{School of Physics, University of Melbourne, Victoria 3010} % Melbourne
% \author{M.~Barrett}\affiliation{High Energy Accelerator Research Organization (KEK), Tsukuba 305-0801} % KEK
% \author{M.~Bauer}\affiliation{Institut f\"ur Experimentelle Teilchenphysik, Karlsruher Institut f\"ur Technologie, 76131 Karlsruhe} % Karlsruhe
  \author{P.~Behera}\affiliation{Indian Institute of Technology Madras, Chennai 600036} % IITM
% \author{C.~Bele\~{n}o}\affiliation{II. Physikalisches Institut, Georg-August-Universit\"at G\"ottingen, 37073 G\"ottingen} % Goettingen
% \author{K.~Belous}\affiliation{Institute for High Energy Physics, Protvino 142281} % Protvino
% ...
% Lebedev
% \author{V.~Zhulanov}\affiliation{Budker Institute of Nuclear Physics SB RAS, Novosibirsk 630090}\affiliation{Novosibirsk State University, Novosibirsk 630090} % BINP
% \author{T.~Zivko}\affiliation{J. Stefan Institute, 1000 Ljubljana} % Ljubljana
% \author{A.~Zupanc}\affiliation{Faculty of Mathematics and Physics, University of Ljubljana, 1000 Ljubljana}\affiliation{J. Stefan Institute, 1000 Ljubljana} % Ljubljana
% \author{N.~Zwahlen}\affiliation{\'Ecole Polytechnique F\'ed\'erale de Lausanne (EPFL), Lausanne 1015} % Lausanne
\collaboration{The Belle Collaboration}